\documentclass[twocolumn,pre]{revtex4}
\usepackage{graphicx}
\usepackage{amssymb}

\newcommand{\be}{\begin{equation}}
\newcommand{\ee}{\end{equation}}

\begin{document}

\title{Spectral matrix methods for partitioning power grids:\\
Applications to the Italian and Floridian high-voltage networks}

\author{Ibrahim Abou Hamad$^a$}
\author{Brett Israels$^{a,b}$}
\author{Per Arne Rikvold$^a$}
\author{Svetlana V.\ Poroseva$^b$}

\affiliation{
$^a$ Department of Physics and Center for Materials 
Research and Technology\\
Florida State University, Tallahassee, FL 32306-4350, U.S.A.\\
$^b$ Center for Advanced Power Systems,  
Florida State University, Tallahassee, FL 32310, U.S.A.
}

\begin{abstract}
\noindent
{\bf Abstract:} 
Intentional islanding is used to limit cascading 
power failures by isolating highly connected ``islands" with 
local generating capacity. To efficiently 
isolate an island, one should break as few power lines as possible. 
This is a {\it graph partitioning\/} problem, and here we give 
preliminary results on islanding of the Italian and Floridian high-voltage 
grids by spectral matrix methods.\\
{~}
\end{abstract}

\maketitle

\section{Introduction}
\label{sec-Int}

Large-scale blackouts have devastating effects on the economy and welfare 
of any modern society \cite{LILI05,PEIR09}. 
One of the reasons \cite{ANDE05,DOBS07} that make 
such catastrophic events possible is the lack of a pre-planned 
strategy for splitting a power grid into separate parts with independent 
generation, also called islands \cite{IEEE07}. 
This defensive strategy, called 
planned, intentional, controlled, or defensive 
islanding, is a last-resort, but effective means to prevent cascading 
outages \cite{PEIR09,YANG06}.   

Intentional islanding splits a power system into islands by breaking 
selected transmission lines. Multiple approaches 
(see, e.g., \cite{LILI05,PEIR09,YANG06,CHOW08,WANG04,LIU06}) 
have been suggested for optimizing the 
selection of the lines to be cut. 
Most analyze the system state, steady 
or dynamic. A useful contribution to these studies can be an analysis 
of the system topology based on a representation of the network 
as a graph \cite{SEAR03,FORT10,NEWM04C,ROSA07}. 
Identification of ``weak" links, whose removal can split 
a given network into independent islands can be beneficial for 
i) initiating fast predetermined intentional islanding and 
ii) preventing unintentional islanding. If one knows in 
advance the minimal set of links that must be broken
to create a separate island, a decision on 
intentional islanding can be made very fast. These 
links should also be closely monitored, as their removal 
(e.g., by accident or sabotage) will result with 
certainty in unintentional islanding. 
The advantages of this strategy are that i) the islanded areas 
can be planned and analyzed in advance with regard to their 
generating capacity and necessary load-shedding if an island has to 
be formed, ii) islands do not depend on the system disturbance and 
coherency of generators \cite{AVRA80}, 
iii) depending on the scale of the event, several  
islands can be formed. Furthermore, this approach is fully compatible 
with other techniques \cite{LILI05,PEIR09,YANG06,CHOW08,WANG04,LIU06}. 

Here we present some preliminary results on using spectral matrix 
methods for intentional islanding of utility power grids, illustrated 
by applications to the Italian and Floridian high-voltage grids. 
The methods are briefly outlined in Sec.~\ref{sec:met}, numerical 
results are presented in Sec.~\ref{sec:res}, and some 
conclusions are drawn in Sec.~\ref{sec:dis}. 

\section{Methods}
\label{sec:met}
\begin{figure}[t] 
\includegraphics[angle=0,width=.496\textwidth]{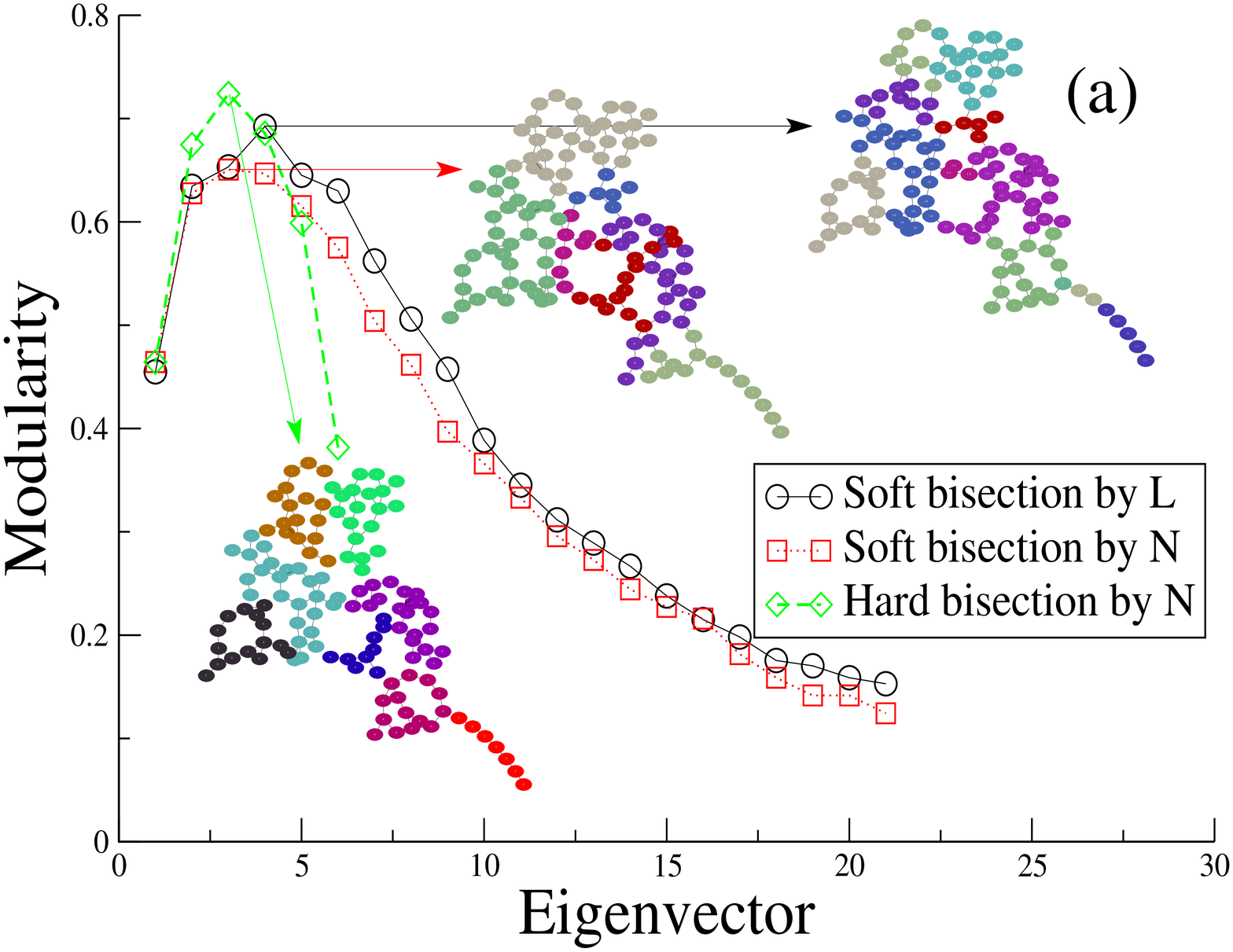} 
\includegraphics[angle=0,width=.496\textwidth]{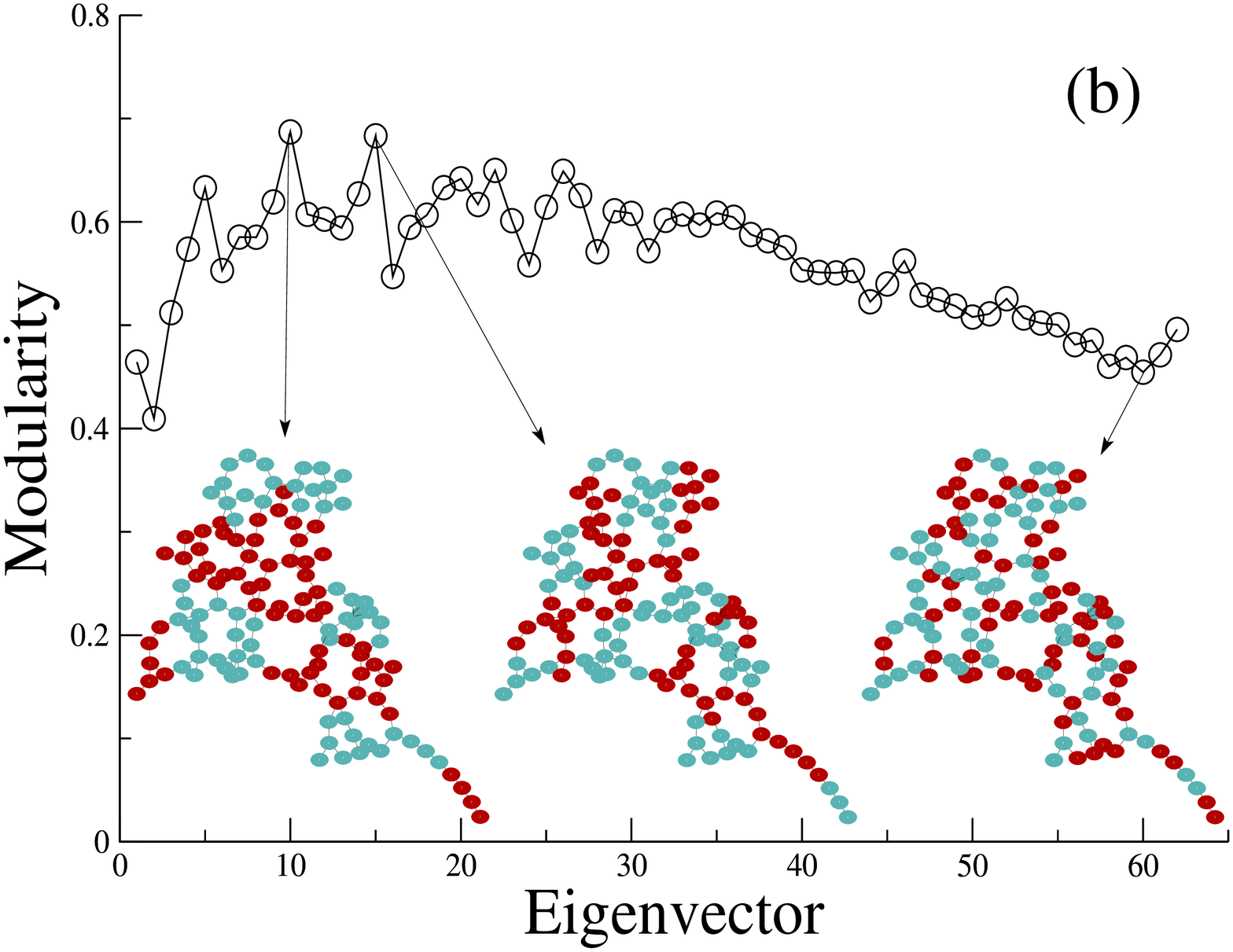} 
\caption[]{
Modularities and islanding configurations for the Italian grid, 
shown vs.\ eigenvector number, as
obtained by bisection methods (a) and from single eigenvectors (b). 
See details in the text.  
}
\label{fig:IT}
\end{figure}

We represent a 
power grid by an undirected graph \cite{SEAR03,NEWM04C}, defined by the 
$N \times N$ symmetric {\it weight matrix\/} $\bf W$, whose elements 
$w_{ij} \ge 0$ represent the capacities of 
the transmission lines (edges) between the $N$ locations 
(vertices) $i$ and $j$. (If all $w_{ij}$ are either 0 or 1, $\bf W$ is 
known as the {\it adjacency matrix\/}.) 
Examples of the graph 
representations of the Italian \cite{ROSA07} and Floridian \cite{FLAMAP} 
high-voltage grids at various levels of islanding are 
shown in Figs.~\ref{fig:IT} and~\ref{fig:FL}, respectively. 

The row sums of $\bf W$, 
$w_i = \sum_j w_{ij}$, are the {\it vertex strengths\/}, 
and $w = \sum_i w_i$ is the {\it total strength\/} of the 
graph. (In the unit-weight case, the $w_i$ are known as vertex
{\it degrees\/}, and $w$ is twice the total number of edges.) The 
list of vertex strengths, $\{w_i\}$, defines a diagonal matrix $\bf D$.  
Spectral graph analysis is usually not performed directly on 
$\bf W$, but rather on one of several matrices derived 
from it. The most common ones are the {\it Laplacian matrix\/} and 
the {\it Normal matrix\/} \cite{SEAR03,FORT10}. 
The Laplacian is defined as 
${\bf L} = {\bf D} - {\bf W}$ and is symmetric 
with vanishing row sums. It embodies Kirchhoff's laws and 
represents a simple resistor network with conductances $w_{ij}$. 
Multiplied with a column vector $| \phi \rangle$ of vertex potentials, 
it yields the vector of currents entering the circuit at each vertex. 
The eigenvalue problem, 
\be
{\bf L} | \phi \rangle = \lambda | \phi \rangle \;,
\label{eq:LL}
\ee
has (at least) one zero eigenvalue, whose eigenvector 
corresponds to equal potentials at each vertex. If the zero eigenvalue is 
$k$-fold degenerate, the graph has $k$ 
disjoint parts. The signs of the components of the eigenvector 
corresponding to the {\it smallest nonzero eigenvalue\/} $\lambda_1$ 
(the Fiedler vector $| \phi_1 \rangle$ \cite{SEAR03}) provide 
a partition of the network into two {\it almost\/} disconnected parts 
(``min-cut theorem" \cite{SEAR03,ROSA07}). 

The normal matrix $\bf N$ is defined by its elements, 
$n_{ij} = w_{ij}/w_i$, so that all its row-sums equal unity (i.e., 
it is a {\it row-stochastic\/} matrix). {\it Left\/} multiplication 
by a vector representing a probability distribution,
$\langle p(t) | {\bf N} = \langle p(t+1) | $, describes 
a discrete-time random walk along the edges of the graph. 
An eigenvalue problem is now given by 
\be
\langle \psi | {\bf N} = \langle \psi | \mu \;.
\label{eq:N}
\ee
The largest eigenvalue of $\bf N$ equals unity, and the corresponding 
left eigenvector (properly normalized) corresponds to the equilibrium 
distribution, $p_i^0 = w_i/w$. (If the unit eigenvalue is $k$-fold 
degenerate, the graph has $k$ disjoint parts.) 
The eigenvector corresponding to the second-largest eigenvalue represents 
the most slowly relaxing perturbation away from the equilibrium 
distribution, and the signs of its components identify two {\it almost\/} 
disconnected sets.
While the Laplacian depends only on the 
off-diagonal part of $\bf W$, the Normal matrix also depends on the 
diagonal terms, $w_{ii}$, which represent {\it self loops\/} in the 
graph that may endow vertices with internal structure. 

The islanding problem is one of partitioning the power grid into 
communities of vertices that are highly interconnected among themselves, 
but only sparsely connected to the rest or the graph. So, ideally 
one would like to find a partitioning into a ``suitable" number of 
communities while maximizing the number of intra-community edges 
and minimizing the number of inter-community edges. This problem is 
NP-hard \cite{FORT10}, 
and so one has to resort to heuristics 
producing ``reasonably good," approximate solutions. 
For the islanding to be useful, each island should contain at least 
one generating plant. 

There are several ways that the eigenvectors of $\bf L$ or 
$\bf N$ can be used to partition a graph in such an approximate 
fashion. Two are based on {\it successive 
bisections\/}. At the first step they are identical: one divides 
the graph into two parts or ``communities" 
according to the signs of the components of the first 
nontrivial eigenvector (i.e., the one with the smallest nonzero 
eigenvalue for $\bf L$, or the one with the eigenvalue closest below 
unity for $\bf N$). In the following steps one can either 
a) continue to successively bisect the network into quadrants, octants, etc., 
according to the signs of the components of the 
next following eigenvectors (``soft bisection"), 
or b) remove the edges connecting the two parts obtained in the first 
step, calculate the first nontrivial eigenvectors of each part 
separately, bisect each according to the signs of their 
respective eigenvector components, and then repeat this procedure with the 
individual parts as often as desired \cite{NEWM04C} (``hard bisection"). 
A third possible method could use each eigenvector separately, labeling each 
vertex $+1$ or $-1$ according to the sign of the corresponding 
eigenvector component, and then identify communities 
with the separate ``Ising clusters" generated by that particular eigenvector.  
 
The quality of a particular partitioning of the graph
into $M$ communities, ${\mathcal C} = \{C_1, ..., C_M\}$ 
can be quantified by Newman's {\it modularity\/} \cite{NEWM04C}. 
It gives the difference between the 
proportion of edges that are internal to a community in the particular 
graph, and the average of the same proportion in a null-model that preserves 
the individual vertex strengths, $w_i$
(and consequently also the total strength, $w$), but is otherwise 
randomly connected.
It is defined as follows:
\be
Q = \frac{1}{w} \sum_{ij} \left( w_{ij} - \frac{w_i w_j}{w}  \right)
\delta \left( C(i),C(j) \right)
\;,
\label{eq:Q}
\ee
where $\delta \left( C(i),C(j) \right) = 1$ if vertices $i$ and $j$ 
are in the same community, and vanishes otherwise. 

\section{Results}
\label{sec:res}

\subsection{Italy}

The Italian 380~kV grid \cite{ROSA07} was 
modeled as an undirected graph of 127 vertices 
and 169 edges of unit weight. We had no information to 
distinguish between vertices representing generating plants, substations, etc.,
so they are all treated as equivalent. International connections are ignored. 
Modularity and partitionings for the Italian grid are shown vs.\ eigenvector 
number in Fig.~\ref{fig:IT}. The results of soft bisection 
based on $\bf L$ 
and $\bf N$ and hard bisection based on $\bf N$ are shown in 
Fig.~\ref{fig:IT}(a), and partitionings based on higher-order, 
single eigenvectors are shown in Fig.~\ref{fig:IT}(b). 

A maximum modularity of $Q \approx 0.72$ was obtained at 
the third level of hard bisection based on $\bf N$ (8 islands with at 
least 7 vertices each). Among the bisection methods, this 
is followed by $Q \approx 0.69$ at the fourth level of soft 
bisection based 
on $\bf L$ (12 islands including one single vertex), and $Q \approx 0.65$ 
at the third level of soft bisection based on $\bf N$ 
(7 ``islands" including two that are internally disconnected). 
The modularities obtained by all the bisection methods 
decrease rapidly beyond the third or fourth level of bisection.

\begin{figure}[t] 
\includegraphics[angle=0,width=.496\textwidth]{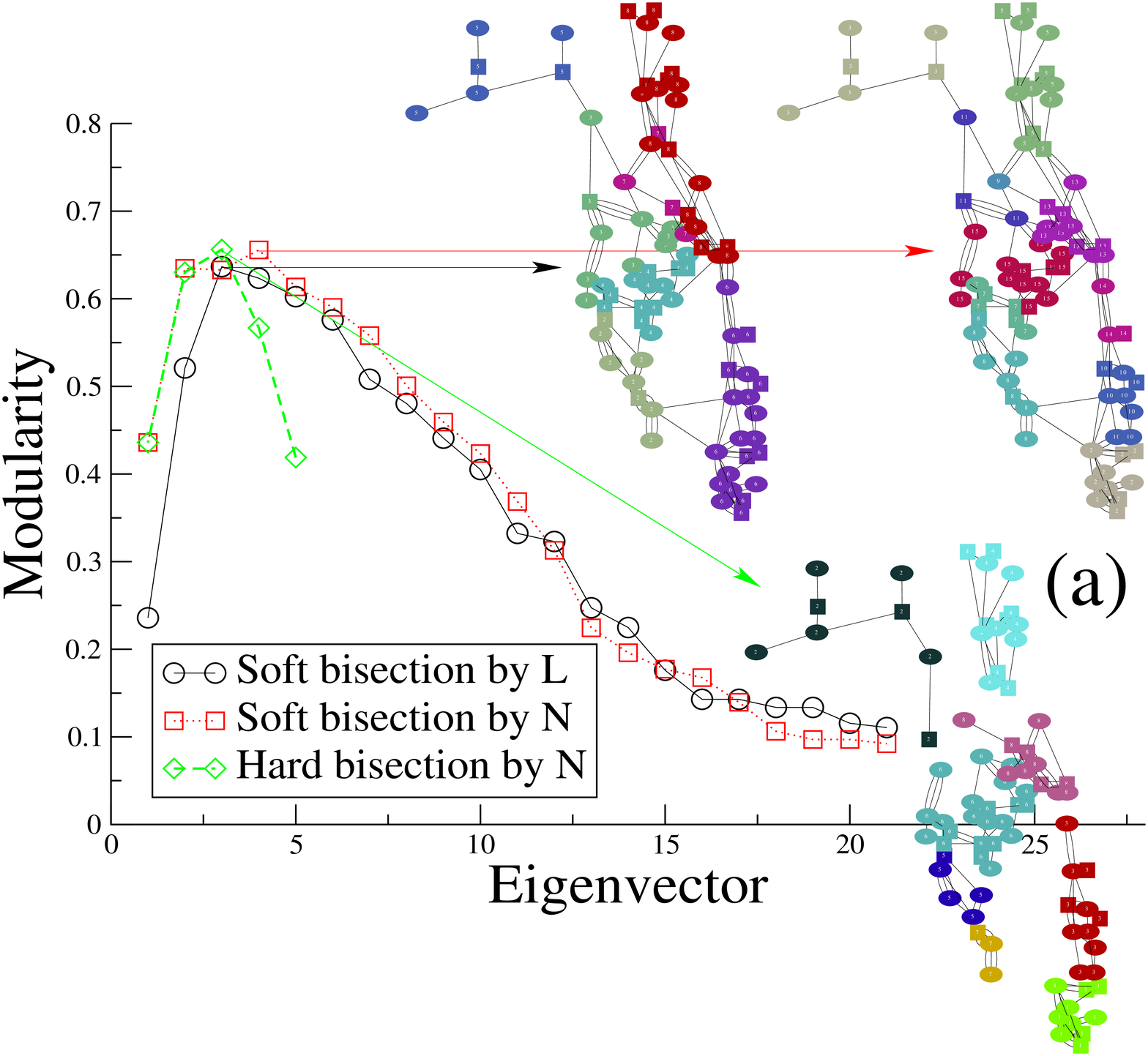} 
\includegraphics[angle=0,width=.496\textwidth]{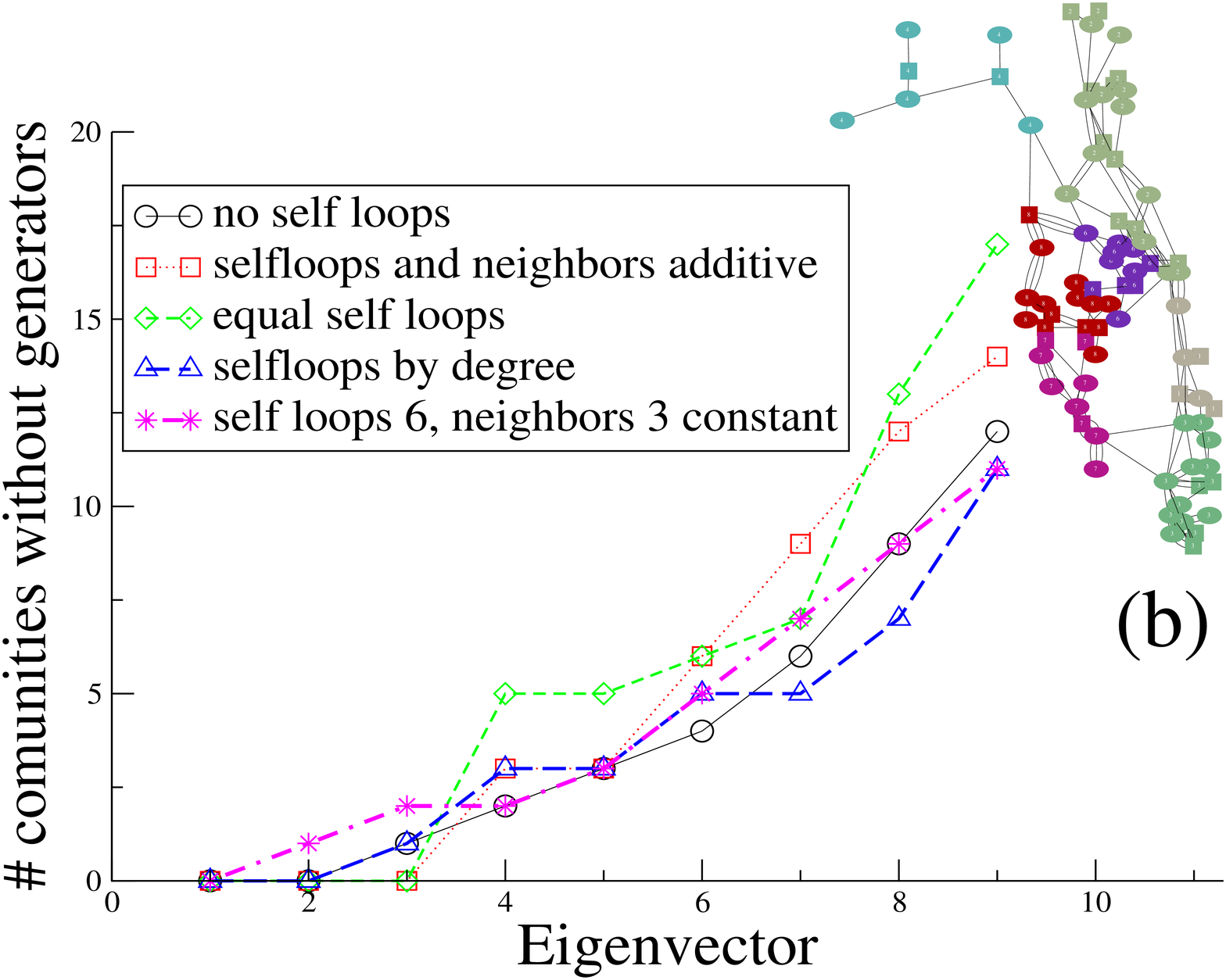} 
\caption[]{Results for the Floridian grid, 
shown vs.\ eigenvector number. 
(a) Modularities and islanding configurations
obtained by bisection with all vertices equivalent. 
(b) 
The number of islands without generators for several schemes 
that distinguish the generators by self loops. Circles: no self loops. 
Diamonds: 6 self loops per generator. 
Triangles: number of self loops equal to generator vertex strength. 
Squares: self loops for generators and their nearest neighbors according 
to strength. 
Star: 6 self loops for generators and 3 for each of their 
nearest neighbors. 
The inset represents the three-level soft bisection with 6 self loops 
per generator. 
See details in the text.  
}
\label{fig:FL}
\end{figure}

The modularities obtained from the ``Ising clusters" defined by individual 
eigenvectors are somewhat irregular and also decrease much more slowly 
with the
number of the eigenvector used, than do the results from the bisection 
methods. The maxima are $Q \approx 0.69$ for the 10th eigenvector 
(8 islands with at least 4 vertices each), 
and $Q \approx 0.68$ for the 15th eigenvector (11 islands with at least 3 
vertices each), respectively. 
The 60th eigenvector still yields a partition with $Q \approx 0.45$, 
but it produces approximately 40 islands. 

\subsection{Florida}

The map of the Floridian high-voltage grid \cite{FLAMAP} is a 
composite of three networks (500, 230, and 138~kV) with 84 vertices, 
31 of which are generating plants. 
We have modeled it as an undirected graph with 
137 edges. The edges have integer weights between 1 and 4, 
according to the actual number of direct lines between pairs of connected 
vertices. Interstate connections are ignored (except for including two 
generating plants and four substations in southern Georgia). 

Modularity and partitionings for the Floridian grid based on bisection 
with all vertices equivalent are shown vs.\ eigenvector number in 
Fig.~\ref{fig:FL}(a). 
Maximum modularities of $Q \approx 0.66$ were obtained at 
the third level of hard bisection based on $\bf N$ (8 islands with at 
least 3 vertices each, all with generating plants) and the fourth 
level of soft 
bisection based on $\bf N$ [11 islands, including one single vertex 
(not a generating plant) and one internally disconnected ``island" in 
which the smaller part has no generator]. This 
is followed by $Q \approx 0.64$ at the third level of soft 
bisection based on $\bf L$ (7 islands with at 
least 6 vertices each, all with generators). 
As for Italy, the modularities decrease rapidly beyond the 
third or fourth level of bisection. 

Bisections according to $\bf N$ offer the opportunity to give
extra weight 
to generating plants by attaching self loops. Several schemes were 
tested, as described in the caption of Fig.~\ref{fig:FL}(b). In some 
cases this enables us to increase the highest level of bisection that  
ensures generators in all islands from two to three. An example for three-level 
soft bisection with six self loops per generator
(7 islands with a minimum of 6 vertices, $Q \approx 0.70$) 
is shown as an inset. 

\section{Conclusions}
\label{sec:dis}

Our results indicate that spectral matrix method can be used, at least to 
obtain an initial partitioning of a power grid into islands. As bisection 
methods (Fig.~\ref{fig:IT}(a) and Fig.~\ref{fig:FL}) 
require evaluation of only a few, dominant eigenvectors, they 
are computationally much more economical than methods based on 
higher-order, single eigenvectors (Fig.~\ref{fig:IT}(b)). 
With the normal matrix method, generating plants can be 
weighted by the introduction of self-loops to increase the probability 
that each island has at least one generator.

\section*{Acknowledgments}
\label{sec:ack}

This work was supported in part by 
U.S.\ National Science Foundation Grant No.\ DMR-0802288, 
U.S.\ Office of Naval Research Grant No.\ 
N00014-08-1-0080, and the Institute for Energy Systems, Economics, 
and Sustainability at Florida State University. 






\begin{thebibliography}{10}
\expandafter\ifx\csname url\endcsname\relax
  \def\url#1{\texttt{#1}}\fi
\expandafter\ifx\csname urlprefix\endcsname\relax\def\urlprefix{URL }\fi
\expandafter\ifx\csname href\endcsname\relax
  \def\href#1#2{#2} \def\path#1{#1}\fi

\bibitem{LILI05}
H.~Li, G.~W. Rosenwald, J.~Jung, C.~Liu, 
Strategic power infrastructure defense, 
Proc.\ IEEE 93 (2005) 918--933.

\bibitem{PEIR09}
A.~Peiravi, R.~Ildarabadi, 
A fast algorithm for intentional islanding of power
  systems using the multilevel kernel $k$-means approach, 
J.\ Appl.\ Sci. 9 (2009) 2247--2255.

\bibitem{ANDE05}
G.~Andersson, 
P.~Donalek, R.~Farmer, N.~Hatziargyriou, I.~Kamwa, P.~Kundur,
  N.~Martins, J.~Paserba, P.~Pourbeik, J.~Sanchez-Gasca, R.~Schultz,
  A.~Stankovic, C.~Taylor, V.~Vittal, 
Causes of the 2003 major grid blackouts
  in North America and Europe and recommended means to improve system 
dynamic performance, 
IEEE Trans.\ Power Syst. 20 (2005) 1922--1928.

\bibitem{DOBS07}
I.~Dobson, B.~A. Carreras, V.~E. Lynch, 
Complex systems analysis of series of blackouts: 
Cascading failure, critical points, and self-organization, 
Chaos 17 (2007) 026103.

\bibitem{IEEE07}
{IEEE Standards Coordinating Committee 21}, IEEE Std.\ 1547.3-2007, 
IEEE Guide
for Monitoring, Information Exchange, and Control of Distributed Resources
  Interconnected with Electric Power Systems, IEEE, 2007.

\bibitem{YANG06}
B.~Yang, V.~Vittal, G.~T. Heydt, 
Slow-coherency-based controlled islanding -- a demonstration of the 
approach on the august 14, 2003 blackout scenario, 
IEEE Trans.\ Power Syst. 21 (2006) 1840--1847.

\bibitem{CHOW08}
S.~P. Chowdhury, S.~Chowdhury, C.~F. Ten, P.~A. Crossley, 
Islanding protection of distribution systems with distributed generators -- 
a comprehensive survey report, 
In: Proceedings of Power and Energy Society General Meeting -
Conversion and Delivery of Electrical Energy in the 21st Century, 
IEEE, 2008.

\bibitem{WANG04}
X.~Wang, V.~Vittal, 
System islanding using minimal cutsets with minimum net flow, 
In: 2004 IEEE PES Power Syst.\ Conf.\ and Exposition, vol.\ 1,
  IEEE, 2004, pp. 379--384.

\bibitem{LIU06}
Y.~Liu, Y.~Liu, 
Aspects on power system islanding for preventing widespread blackout, 
In: Proceedings of the 2006 IEEE International Conference on
  Networking, Sensing and Control, ICNSC'06, Ft. Lauderdale, FL, 
IEEE, 2006, pp. 1090--1095.

\bibitem{SEAR03}
A.~J. Seary, W.~D. Richards, 
Spectral methods for analyzing and visualizing networks: 
An introduction, 
In: R.~A. Breiger (Ed.), Dynamic Social Network
  Modeling and Analysis, National Academies' Press, Washington, DC, 2003.

\bibitem{FORT10}
S.~Fortunato, 
Community detection in graphs, 
Phys.\ Rep. 486 (2010) 75--174, and references cited therein.

\bibitem{NEWM04C}
M.~E.~J. Newman, 
Analysis of weighted networks, 
Phys.\ Rev.\ E 70 (2004) 056131.

\bibitem{ROSA07}
V.~Rosato, S.~Bologna, F.~Tiriticco, 
Topological properties of high-voltage electrical transmission networks, 
El.\ Power Syst.\ Res. 77 (2007) 99--105.
  Table 1 gives the number of edges of the Italian grid as 171,
  but we have used only the 169 that we could discern from the 
map in Fig.~3 of this paper.

\bibitem{AVRA80}
B.~Avramovic, P.~V. Kokotovic, J.~R. Winkelman, J.~H. Chow, 
Area decomposition for electromechanical models of power systems, 
Automatica 16 (1980) 637--648.

\bibitem{FLAMAP}
S.~Dale, 
T.~Alquthami, T.~Baldwin, O.~Faruque, J.~Langston, P.~McLaren,
  R.~Meeker, M.~Steurer, K.~Schoder, 
Progress Report for the Institute for Energy Systems, 
Economics and Sustainability and the Florida Energy Systems
  Consortium, Florida State University, Tallahassee, FL, 2009.

\end{thebibliography}







\end{document}